# Bound states in and out of the continuum in nanoribbons with wider sections: A novel recursive S-matrix method

Ricardo Y. Díaz and Carlos Ramírez

Departamento de Física, Facultad de Ciencias, Universidad Nacional Autónoma de México, Apartado Postal 70542, 04510 Ciudad de México, México

**Abstract**
We report a novel method to find bound states in general tight-binding Hamiltonians with semi-infinite leads. The method is based on the recursive S-matrix method, which allows us to compute iteratively the S-matrix of a general system in terms of the S-matrices of its subsystems. We establish the condition that the S-matrices of the subsystems must accomplish to have a bound state at energy E. Energies that accomplish this relation, can be determined with high accuracy and efficiency by using the Taylor series of the S-matrices. The method allows us to find bound states energies and wavefunctions in (BIC) and out (BOC) of the continuum, including degenerate ones. Bound states in nanoribbons with wider sections are computed for square and honeycomb lattices. Using this method, we verify the bound states in a graphene nanoribbon with two quantum-dot-like structures which has been reported to have BICs by using another technique. However, this new analysis reveals that such BICs are double, one with even and the other with odd wavefunction, with slightly separated energies. In this way, the new method can be used to efficiently find new BICs and to improve precision in previously reported ones.

## 1. Introduction

Bound states in a quantum system have normalizable wavefunctions that are non-zero only in a finite region of the space or decay to zero when the position goes to infinity. For finite closed systems, all wavefunctions are bound states, with discrete energies that can be calculated by diagonalizing the Hamiltonian. Let us consider a general tight binding Hamiltonian with $N$ sites,

$$\hat{H} = \sum_{n=1}^{N} \varepsilon_n |n\rangle\langle n| + \sum_{\substack{n,m=1 \\ n \neq m}}^{N} t_{nm} |n\rangle\langle m| , \qquad (1)$$

where $|n\rangle$ is the Wannier function at site $n$ with site-energy $\varepsilon_n$, and $t_{n,m}$ is the hopping integral between sites $n$ and $m$. The energy $E_m$ of Hamiltonian with eigenfunction

$$|\psi_m\rangle = \sum_{n=1}^{N} a_{n,m} |n\rangle \qquad (2)$$

solves the matrix equation

$$\begin{pmatrix} \varepsilon_1 & t_{1,2} & \cdots & t_{1,N} \\ t_{2,1} & \varepsilon_2 & \cdots & t_{2,N} \\ \vdots & \vdots & \ddots & \vdots \\ t_{N,1} & t_{N,2} & \cdots & t_{N,N} \end{pmatrix} \begin{pmatrix} a_{m,1} \\ a_{m,2} \\ \vdots \\ a_{m,N} \end{pmatrix} = E_m \begin{pmatrix} a_{m,1} \\ a_{m,2} \\ \vdots \\ a_{m,N} \end{pmatrix} \qquad (3)$$

The matrix in Eq. (3) is the matrix representation of the Hamiltonian in the base of the Wannier functions $\{|n\rangle\}$. In other words, by calculating the eigenvalues and eigenvectors of the Hamiltonian matrix, we find the $N$ energies and eigenfunctions of the system. Numerically, this calculation can be done by using standard procedures for diagonalization, such as those in the Math Kernel Library (MKL). However, numerical diagonalization is computationally expensive in large systems, as occurs in systems with leads.

A lead is a semi-infinite structure formed by the periodical repetition of a unitary cell in one direction, while the other side of the lead is connected to the scattering region. Leads act as waveguides for incoming and outgoing Bloch waves (open channels), but they may also contain evanescent functions that decay exponentially to zero away from the scattering region (closed channels). The scattering matrix (S-matrix) of the system relates the coefficients of incoming Bloch waves to those of outgoing Bloch waves. Consequently, the dimension of the S-matrix is equal to the number of open channels in the leads. Due to the periodic nature of the leads, energies associated with Bloch waves form continuum energy bands. Bound states do not overlap with any of the Bloch waves, because wavefunctions associated with Bloch waves are non-normalizable, but they can couple to evanescent modes. Bound states in the continuum (BIC) have energies at values where there are open channels, so they coexist with the energy-bands in leads. On the other hand, bound states out of the continuum (BOC) have energies at values with non-open channels.

The S-matrix of a system contains all the effects caused by the scattering region on its surroundings. It can be found by direct inversion, which could be done efficiently in sparse systems [1]. Also, the S-matrix can be calculated from the S-matrices of their subsystems by using the recursive scattering matrix method (RSMM) [2,3]. By starting from the S-matrix of site and bond structures [2], the matrix of a general tight-binding system with general leads can be found. Moreover, the RSMM has been formulated to obtain the order-N Taylor expansion of the S-matrix [4]. The Landauer formula allows us to use of the S-matrix to calculate the conductance of mesoscopic systems [5]. The S-matrix is also useful to efficiently find the global density of states [6,7], Fano factor [8], Wigner time-delay [9], band structure in quasi-one-dimensional systems [3], and other related physical quantities [10–12]. Recently, a new method based on the S-matrix has been developed to find accurately and efficiently the bound states of the one-dimensional Schrödinger equation for arbitrary potential wells [13].

Being a general wave phenomenon, BICs have been studied in a variety of systems such as optical [14–20], acoustic [21,22] and solid state physics [23–25]. Even though BICs were first proposed by Wigner in 1929 [26], in recent years they have been experimentally observed in diverse optical [15,16,19,20] and solid-state systems [23]. BICs have also been proposed for technological applications such as BIC-based lasers, sensors and waveguides [14,27–30].

Different methods have been developed to determine the energies for which BICs can be found in tight binding systems. For instance, a method was proposed that consists in solving an equation of eigenvalues in terms of the self-energy and associating the null eigenvalues to states decoupled from the extended states [31]. A different method consists in the localization of peaks in the density of states (DOS) that are not accompanied by a peak in the conductance [24]. Alternatively, a compact system may present dissipation proper of an open system by considering a non-Hermitian Hamiltonian. This technique has allowed to find corner BICs in topological insulators for the energies that eliminate the dissipation [25].

In this paper, we present a novel method based on RSMM to find bound states in (BIC) and out (BOC) of the continuum in general tight-binding systems with leads. In section 2, we determine the condition that S-matrices of subsystems must accomplish to find a bound state and explain the method to find its associated wavefunction. Section 3 presents examples of bound states calculated by using this method in square-lattice nanoribbons with wider sections, while Section 4 shows cases with honeycomb-lattice nanoribbons.

## 2. The method

Let us consider a general tight-binding system A with attached semi-infinite atomic chains (auxiliary chains). These auxiliary chains have sites with null site-energy and are connected between first neighbors by hopping integrals $t_C$. Due to the periodicity of auxiliary chains, the amplitude coefficient of the $m$-th site in the $n$-th auxiliary chain $(a_{n,m})$ is given as a linear combination of an incoming and an outgoing wave [2],

$$a_{n,m} = A_n^{(+)} e^{-ikm} + A_n^{(-)} e^{ikm}, \tag{4}$$

where $A_n^{(+)}$ and $A_n^{(-)}$ are respectively the coefficients of the incoming and the outgoing waves in the $n$-th auxiliary chain, and $k$ is related to the energy by $E = 2t_C \cos(k)$. The case $m=0$ corresponds to the site where the auxiliary chain is attached.

Let us consider another system B with auxiliary chains and denote the coefficients of the incoming and outgoing waves in these chains as $B_n^{(+)}$ and $B_n^{(-)}$, respectively. According to the RSMM [2], doing $B_n^{(\pm)} = A_n^{(\mp)}$ for $n = 1, 2, \cdots, N$ allows to model the system that results from the joining of systems A and B, as exemplified in Fig. 1. Notice that these equalities remove the corresponding auxiliary chains. This fact can be used to compute recursively the S-matrix of any system from the S-matrix of its subsystems. By starting from the site and bond structures, which have analytical S-matrices [2], any tight-binding structure can be modeled by following the RSMM, including multiterminal cases with general leads [3]. Moreover, the RSMM has been extended to find the Taylor series of the S-matrix to arbitrary order [4]. The S-matrix of a system is a function of energy, then its Taylor series allows us to compute the exact derivatives of the S-matrix with respect to the energy.

2.1 Determination of energies

Let us assume that we want to find the bound states of a general system C with leads, as the one shown in Fig. 1(a). Due to the leads, the number of sites in system C is infinite, and then the bound states cannot be exactly solved by numerical direct diagonalization. Following the RSMM, we can divide system C into two subsystems A and B, as shown in Fig. 1(b). Sites in the frontier of this division are connected to auxiliary chains. In this way, scattering matrices of systems A and B can be calculated independently as functions of the energy. S-matrix of system A satisfy

$$\begin{pmatrix} \mathbf{A}^{(-)} \\ \mathbf{L}^{(-)} \end{pmatrix} = \begin{pmatrix} \mathbf{S}_A & \mathbf{S}_{AL} \\ \mathbf{S}_{LA} & \mathbf{S}_{LL} \end{pmatrix} \begin{pmatrix} \mathbf{A}^{(+)} \\ \mathbf{L}^{(+)} \end{pmatrix}. \tag{5}$$

where

$$\mathbf{A}^{(\pm)} = \begin{pmatrix} A_1^{(\pm)} \\ A_2^{(\pm)} \\ \vdots \\ A_N^{(\pm)} \end{pmatrix}, \quad \mathbf{L}^{(\pm)} = \begin{pmatrix} L_1^{(\pm)} \\ L_2^{(\pm)} \\ \vdots \\ L_N^{(\pm)} \end{pmatrix}, \tag{6}$$

with coefficients $L_n^{(\pm)}$ associated with incoming (+) and outgoing (-) Bloch functions (open channels) in the leads. Since bound states should not couple to open channels, we are only interested in cases where $\mathbf{L}^{(\pm)} = 0$. Let us assume that for $\mathbf{A}^{(\pm)} = \mathbf{A}_i^{(\pm)}$ this occurs, then Eq. (5) implies

$$\mathbf{S}_{LA} \mathbf{A}_i^{(+)} = 0, \tag{7}$$

and

$$\mathbf{A}_i^{(-)} = \mathbf{S}_A \mathbf{A}_i^{(+)}. \tag{8}$$

Eq. (7) is accomplished if the incoming wave from the auxiliary chains is fully reflected into the auxiliary chains, which occurs when

$$\left\| \mathbf{A}_i^{(-)} \right\| = \left\| \mathbf{A}_i^{(+)} \right\|. \tag{9}$$

An analogous procedure may be followed for system B to obtain

$$\mathbf{B}_i^{(-)} = \mathbf{S}_B \mathbf{B}_i^{(-)}. \tag{10}$$

Since system C is recovered by doing $\mathbf{A}^{(\pm)} = \mathbf{B}^{(\mp)}$, we conclude that $\mathbf{A}_i^{(+)}$ corresponds to a bound state of system C if it accomplishes Eqs. (8) and (9), and

$$\mathbf{A}_i^{(+)} = \mathbf{S}_B \mathbf{A}_i^{(-)}, \tag{11}$$

where Eq. (10) was used. Since S-matrix is unitary, then Eq. (9) is satisfied whenever Eqs. (8) and (11) are satisfied. On the other hand, combining Eqs. (8) and (11) lead us to

$$\mathbf{A}_i^{(+)} = \mathbf{S}_B \mathbf{S}_A \mathbf{A}_i^{(+)}. \tag{12}$$

Consequently, there is a bound state at energy E whensoever $E$ has an eigenvalue one. Observe that in one-dimensional systems this condition is in agreement with the one reported in Ref. [13].

The algorithm to determine numerically the bound state energies consists of calculating the eigenvalues $\lambda_i$ of the matrix product $\mathbf{S}_A \mathbf{S}_B$ for different energies within an interval of interest, obtaining a set of curves for the imaginary and real parts of the eigenvalues. For example, Fig. 1(d) shows the real and imaginary parts of the eigenvalues of $\mathbf{S}_A \mathbf{S}_B$ for the system in Fig. 1d. In this case, we find bound states at energies $E_1 = 2.36481|t|$ and $E_2 = 2.82856|t|$.

To optimize the searching of bound states, we can use the procedures explained in Ref. [4] to find the order-N Taylor series of each eigenvalue of $\mathbf{S}_A\mathbf{S}_B$,

$$\lambda_i(E_0 + \Delta E) = \sum_{n=0}^{N} c_{i,n}(\Delta E)^n. \tag{13}$$

Exact derivatives of $\lambda_i(E)$ are then given by $\lambda_i^{(n)}(E_0) = n!c_{i,n}$. In this way, we may use the Newton-Raphson method to find bound states with high accuracy and efficiency by looking for the zeroes of the imaginary-part of any of the eigenvalues $\lambda_i(E)$. Once a zero of the imaginary-part is located at energy $E$, we must confirm that the corresponding real-part is one to conclude that we have found a bound state.

2.2 Determination of wavefunctions

Once a bound state energy is found, the corresponding eigenfunction can also be calculated by using the RSMM.

Firstly, according to Eq. (4) we have $\mathbf{a}_i = \mathbf{A}_i^{(+)} + \mathbf{A}_i^{(-)}$, where $\mathbf{a}_i$ is a vector that contains the amplitude coefficients of sites connected to auxiliary chains (in Eq. (4), those with $m=0$) and $\mathbf{A}_i^{(+)}$ is the eigenvector of $\mathbf{S}_A\mathbf{S}_B$ with eigenvalue one, as stated in Eq. (12). By using Eq. (8), we have

$$\mathbf{a}_i = (\mathbf{I} + \mathbf{S}_A)\mathbf{A}_i^{(+)}. \tag{14}$$

This allows us to find amplitude coefficients for the sites shared by subsystems A and B. Now, let us divide system A into two subsystems A' and B' as exemplified in Fig. 1(c). Then, by analogy, coefficients at the frontier of this division can be calculated from

$$\mathbf{a}'_i = (\mathbf{I} + \mathbf{S}_{A'})\mathbf{A}_i^{\prime(+)}. \tag{15}$$

Scattering matrices of systems A' and B' satisfy respectively

$$\begin{pmatrix} \mathbf{A}^{\prime(-)} \\ \mathbf{L}^{(-)} \end{pmatrix} = \begin{pmatrix} \mathbf{S}_{A'} & \mathbf{S}_{A'L} \\ \mathbf{S}_{LA'} & \mathbf{S}_{L'} \end{pmatrix} \begin{pmatrix} \mathbf{A}^{\prime(+)} \\ \mathbf{L}^{(+)} \end{pmatrix}. \tag{16}$$

and

$$\begin{pmatrix} \mathbf{A}^{(-)} \\ \mathbf{B}^{\prime(-)} \end{pmatrix} = \begin{pmatrix} \mathbf{S}_{AA} & \mathbf{S}_{AB'} \\ \mathbf{S}_{B'A} & \mathbf{S}_{B'B'} \end{pmatrix} \begin{pmatrix} \mathbf{A}^{(+)} \\ \mathbf{B}^{\prime(+)} \end{pmatrix} \tag{17}$$

Using the RSMM, system A is recovered by taking

$$\mathbf{A}^{\prime(\pm)} = \mathbf{B}^{\prime(\mp)}. \tag{18}$$

In this way, amplitude coefficients for sites in the frontier between A' and B' are given by $\mathbf{a}'_i = \mathbf{A}_i^{\prime(+)} + \mathbf{A}_i^{\prime(-)}$. Since we are solving the wavefunction for the case of $\mathbf{A}^{(+)} = \mathbf{A}_i^{(+)}$, then we do not have overlaps with Bloch modes in the leads, i.e., $\mathbf{L}^{(\pm)} = 0$. Consequently, Eq. (16) leads to

$$\mathbf{A}_i^{\prime(-)} = \mathbf{S}_{A'}\mathbf{A}_i^{\prime(+)}. \tag{19}$$

On the other hand, Eqs. (17) and (18) give us

$$\mathbf{A}_i^{\prime(+)} = \mathbf{S}_{B'A}\mathbf{A}_i^{(+)} + \mathbf{S}_{B'B'}\mathbf{A}_i^{\prime(-)}. \tag{20}$$

Combining equations (15), (19) and (20), we obtain

$$\mathbf{a}'_i = (\mathbf{I}+\mathbf{S}_{A'})(\mathbf{I}-\mathbf{S}_{B'B'}\mathbf{S}_{A'})^{-1}\mathbf{S}_{B'A}\mathbf{A}_i^{(+)}. \tag{21}$$

Notice that this expression can always be evaluated, even if the number of frontier sites in the left-side of B' is different to that in the right-side.

Subsystems A' and B' can be redefined to have different sites at the frontier. In this way, we can obtain all the amplitude coefficients for sites in A. An analogous procedure can be followed to evaluate the amplitude coefficients for sites in B.

## 2.3 Other forms of division

There are many ways to divide any system C into two subsystems A and B. For example, in Fig. 2, we propose two alternative definitions of systems A and B. In both cases, system A contains all the sites of the system C but has coupled (a) ten and (b) one auxiliary chains. By joining systems A and B through the RSMM, these coupled chains are removed, so we obtain the same system of Fig. 1(a) in both cases. Real and imaginary parts of eigenvalues are shown below each instance. The number of eigenvalues is equal to the number of auxiliary chains coupled to system A. Consequently, we can define a case with only one eigenvalue, as occurs in Fig. 2(b), which may be more convenient to analyze. However, it is worth mentioning that for systems with many sites in the dispersion region, eigenvalue curves are smoother in cases with more auxiliary chains, which is desirable when using the Newton-Raphson method. Moreover, cases with a single eigenvalue cannot resolve if the energy has degeneracy, but this can be achieved by changing the form of division to a case with more eigenvalues every time the bound state condition is satisfied.

On the other hand, once a bound state energy is found, the determination of the wavefunction can be optimized by adding auxiliary chains to system A', because in this way such sites become frontier sites and Eq. (21) can be used to obtain their amplitude coefficients at once.

## 3. Bound states in square-lattice nanoribbons

In the following, we present bound states in square-lattice nanoribbons with a wider section calculated by using the method in section 2. In all cases, we consider null site-energies and hopping integral between nearest neighbors $t<0$.

For the system of Fig. 1(a) with $N=50$, $N_L=40$ and $M=100$, the energies of the first five bound states, shown in Fig. 3 are (a) −3.9956984703, (b) −3.9944369231, (c) −3.9841130904, (d) −3.9819853501 and (e) −3.9787213144 in units of $|t|$. Leads in this system have at least one open channel for energies between $-2\left[1+\cos\left(\frac{\pi}{N_L+1}\right)\right]|t|$ and $2\left[1+\cos\left(\frac{\pi}{N_L+1}\right)\right]|t|$, i.e., between energies −3.9941316 and 3.9941316 for this instance. Consequently, energies (a) and (b) are BOCs while (c)-(e) are BICs. All these eigenvalues are non-degenerate, and since Hamiltonian is symmetrical under horizontal and vertical reflections, these wavefunctions are necessarily symmetric or antisymmetric under these reflections. Energy (a) is the ground state and, as expected, its wavefunction does not have nodes. All other wavefunctions have line nodes, whose number increases for bound states with greater energies. Also observe that amplitude in the leads is not zero for all eigenfunctions, but they decay in the leads.

For the case of $N=13$, $N_L=7$ and $M=27$, Fig. 4(a) shows six BICs at the same energy $E=0$. Figure 4(b) illustrates the wavefunctions of two of these BICs. Since these bound states are degenerate, eigenvectors $\mathbf{A}_i^{(+)}$ in Eq. (12) were orthogonalized by using the Gramm-Schmidt algorithm, leading us to orthogonalized wavefunctions. Due to degeneracy, symmetries under horizontal and vertical reflections in the wavefunctions are not guaranteed. There are different values of $N$, $N_L$ and $M$ for which the system of Fig. 1(a) has BICs at $E=0$. Fig. 4(c) illustrate them by varying the values of $M$ and $N_L$ for fixed $N=38$ and $N=39$, where the color scale represents the number of BICs at $E=0$. For cases with $M=j(N+1)-1$, where $j=1,2,3,\cdots$, there are $N-N_L$ BICs at $E=0$. Wavefunctions in these cases have vertical nodal lines every $N+1$ sites. This can be observed in the examples of Fig. 4(b), where all wavefunctions have a vertical nodal line at the center of the system. In general, if $m=(N-i+1)/i$ is an integer for some integer $i$, there are BICs at $E=0$ whenever $M=j(m+1)-1$ and $M>N_L$. It is important to mention that all wavefunctions associated to BICs at $E=0$ are null in the leads, which is expected, because at this energy there are not evanescent modes, and so any bound state cannot extend into the leads. However, disorder would couple the wavefunctions to open channels in the leads, making these BICs metastable.

## 4. Bound states in graphene nanoribbons

In the following, we determine bound states in armchair graphene nanoribbons (AGNR). The first case considered is a two-quantum-dot-like junction in an AGNR shown in Fig. 5(a). Hydrogen passivation was considered by taking the hopping parameters of the dimers at the border of the AGNR as $t_{edge}=1.12t$ (green bonds in Fig. 5(a)), with $t$ being the parameter in all other cases [32]. This system was previously explored by Gonzalez *et. al.* in search of BICs with a method that consisted in determining energies for which there is a peak in the DOS but not in conductance. Three different BICs were reported with that method for energies 0.05, 0.46 and 0.62 in units of $|t|$ [24]. For the same system, the method hereby proposed allowed us to find six different bound states for positive energies at 0.050492, 0.050633, 0.455494, 0.456079, 0.619317 and 0.619912. By rounding these energies to two decimal places, they can be grouped into pairs, each pair matching those previously reported. The six wave functions were calculated and are presented in Fig. 5(b). Observe that the wave functions within a BIC-pair only differ in their symmetry under vertical reflections, one with even and the other with odd wavefunction, and both contribute to the same charge distribution reported in Ref. [24].

The second AGNR explored has a central region formed by a series of continuous expansions and contractions in the width of the ribbon. There are thirteen different bound state energies. Their wavefunctions are presented in Fig. 6. Except for the two with the lowest energies and the two closest to zero, all the cases have at least one open channel and therefore are BICs. The eigenfunctions have no common behavior regarding whether they are totally confined to the wider sections or if they decay in the leads, showing robustness of the method under different circumstances. Also, observe that the function corresponding to the lowest energy has no nodes, as expected for a ground state, and the number of nodes increases as the energy grows. AGNRs with continuous width variation as the ones shown in Fig. 6 have been explored by Gröning *et. al.* [23], who experimentally obtained conductance maps that are in great agreement with the tight binding simulated charge density. Thus, experimental observation of the bound states theoretically predicted here may be possible using techniques as the one abovementioned.

## 5. Discussion and Conclusions

In this paper, we have presented a new method based on the RSMM to calculate energies and wavefunctions of bound states in general tight-binding systems with infinite leads. The method consists in dividing the system into two subsystems. Then we calculate eigenvalues of a matrix $\mathbf{S}_B \mathbf{S}_A$ (12) that is given in terms of the S-matrices of such subsystems. Whensoever an eigenvalue becomes one, there is a bound state. We have used this method to determine BICs and BOCs in square- and honeycomb-lattice nanoribbons with wider sections. We observed that the presence of BICs is strongly dependent on the geometry of the system.

The number of eigenvalues to be analyzed in search for bound states can be controlled as there are multiple ways to define the subsystems. They only must satisfy that by joining them through the RSMM the system is recovered. The number of eigenvalues depends on the number of auxiliary chains in each subsystem. In this way, we have proposed alternatives with different number of eigenvalues. By analyzing the eigenvalues in each alternative, we show that any of them is useful to find the bound states. In particular, we have proposed in Fig. 2(b) an alternative where one of the subsystems have only one site, leading us to a single eigenvalue of $\mathbf{S}_B \mathbf{S}_A$, which is convenient. However, we pointed out that in cases with more auxiliary chains, curves of eigenvalues are smoother, which is also desirable. Moreover, when the wavefunctions is being calculated, it is better to maximize the number of sites with attached auxiliary chains. In practice, we can change the way the subsystems are defined to explode the advantages of each definition.

The method allows us to compute bound-state energies with high precision and can be used together with other methods to determine energy of BICs. Machine precision is achieved in a few iterations following the Newton-Raphson method, where exact first derivatives of the eigenvalues of $\mathbf{S}_B \mathbf{S}_A$ are computed by using Taylor series of the S-matrices [4]. On the other hand, we can explore the occurrence of bound states in a selected domain of interest, where other methods have reported BICs. For example, BICs in Fig. 5 were previously reported by Gonzalez *et. al.* [24], where they identified peaks in the density of states that are not present in the transmission spectrum as signals of BICs. Energy of these BICs were reported with two decimal places. By exploring the vicinity around such energies with the method hereby proposed we confirm these BICs. But actually, there are two BICs for each reported energy. The energy of each pair differs by the 4th decimal digit. Their corresponding wavefunctions have the same absolute value, which make them indistinguishable when the charge distribution is analyzed through the Green's function. Even when the method reported here can work independently to find BICs, we believe that the technique that occupies the density of states and the transmission spectrum is useful to find regions of interest by inspection. Then, the method proposed here may focus on these regions to find energies and wavefunctions of the bound states with high precision.

We expect that the predicted BICs in this paper could be verified in experimental setups. In particular, those in Fig. 6 correspond to a system that has been realized recently to study conductance properties [23]. Other cases may be explored in acoustic or optical systems [20–22]. Simulations of tight-binding Hamiltonians using ultracold atoms in an optical lattice constitute another possible way to observe BICs, as these techniques are in current development and have already allowed to explore diverse quantum phenomena [33–35]. It is worth to mention that this method does not add any additional approximations beyond the tight-binding approach. Moreover, the method can also be considered for continuous systems where a discretization process results in a tight-binding effective model.

Even when we have applied the method to nanoribbons with wider sections, it can also be used on systems with other kind defects, such as impurities, vacancies, dislocations, etcetera, including point, line and planar defects. This method could also be useful to study the formation of charge islands in graphene nanoribbons with edge disorder [36]. Analysis of these scenarios are currently under development.

**Acknowledgments**

This work has been supported by UNAM-DGAPA-PAPIIT IN116819. Computations were performed at Miztli under project LANCAD-UNAM-DGTIC-329

**Figures**

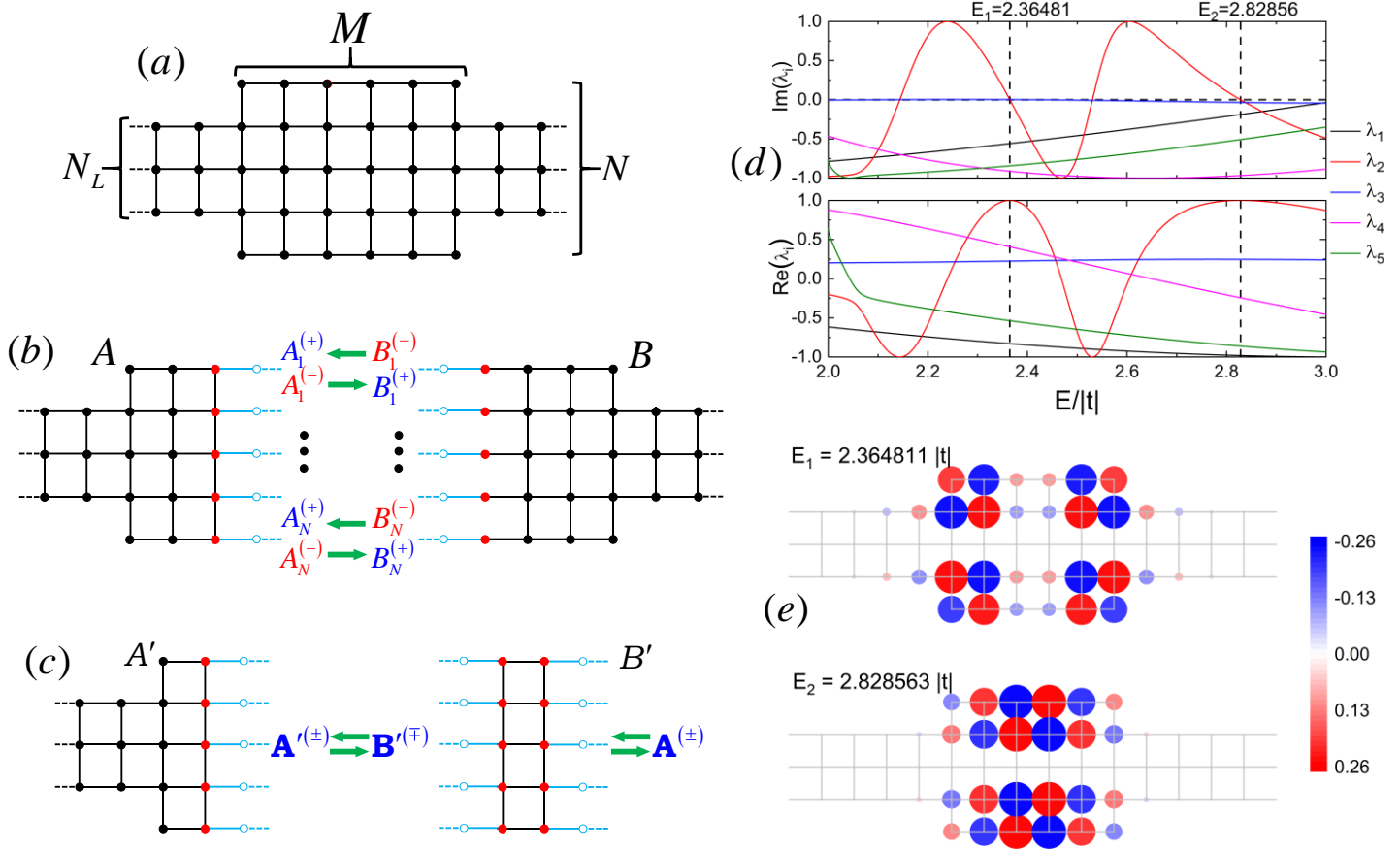

Fig. 1. (a) System C: An infinite nanoribbon with a wider section. An instance with $N_L = 3$, $N = 5$ and $M = 6$ sites is shown. (b) Joining systems A and B through the RSMM we can model system C. Auxiliary chains are represented by blue lines. (c) Joining systems A' and B' through the RSMM, we obtain system A. (d) Eigenvalues of $\mathbf{S}_B \mathbf{S}_A$ for energies between $E = 2|t|$ and $E = 3|t|$. Dashed lines identify energies where the condition for bound states is satisfied by one of the eigenvalues. (d) Wavefunctions associated to the bound-state energies in color and radius scales.

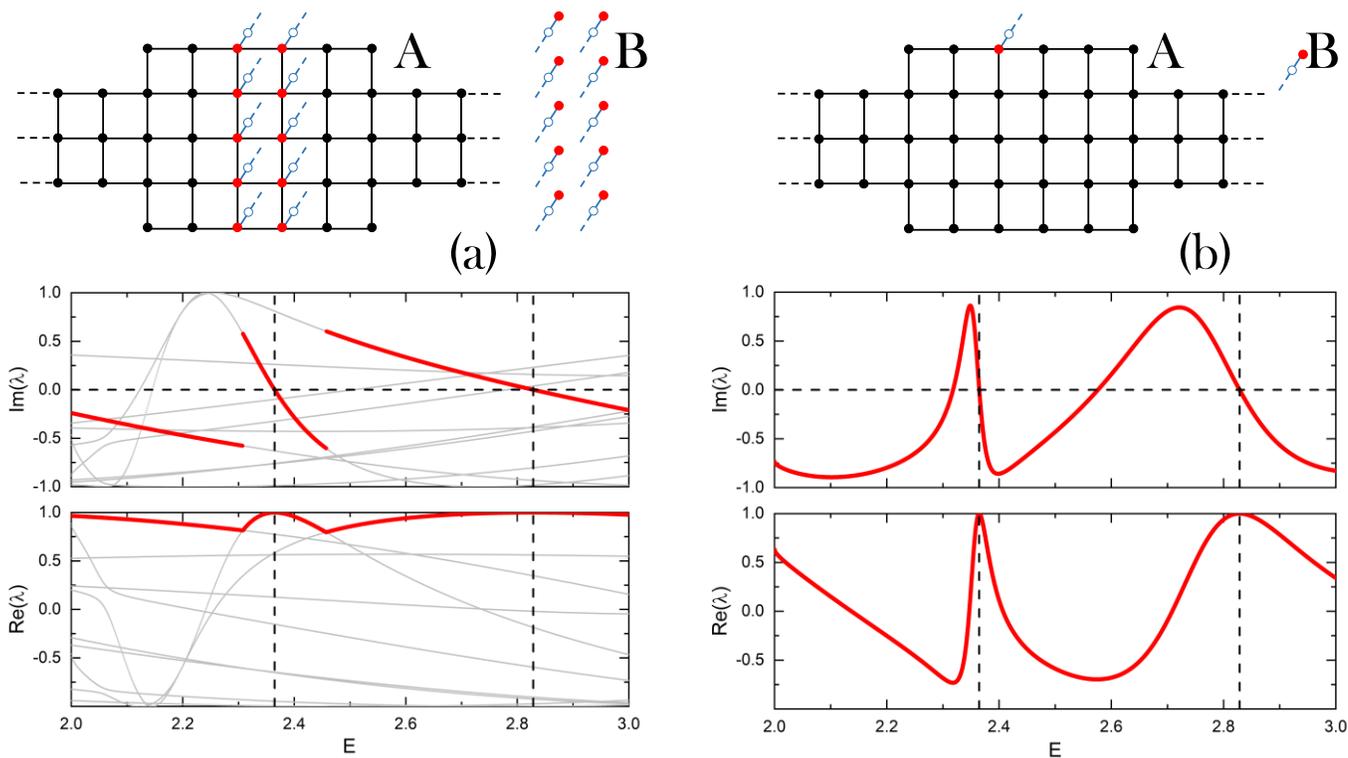

Fig. 2. (top) Two alternative forms of defining systems A and B whose joining through the RSMM lead us to the same system of Fig. 1(a), and (bottom) the corresponding real and imaginary parts of the eigenvalues of $\mathbf{S}_B\mathbf{S}_A$ in each case. The eigenvalue with the greatest real part is highlighted with a thick red line, while other eigenvalues are shown in gray.

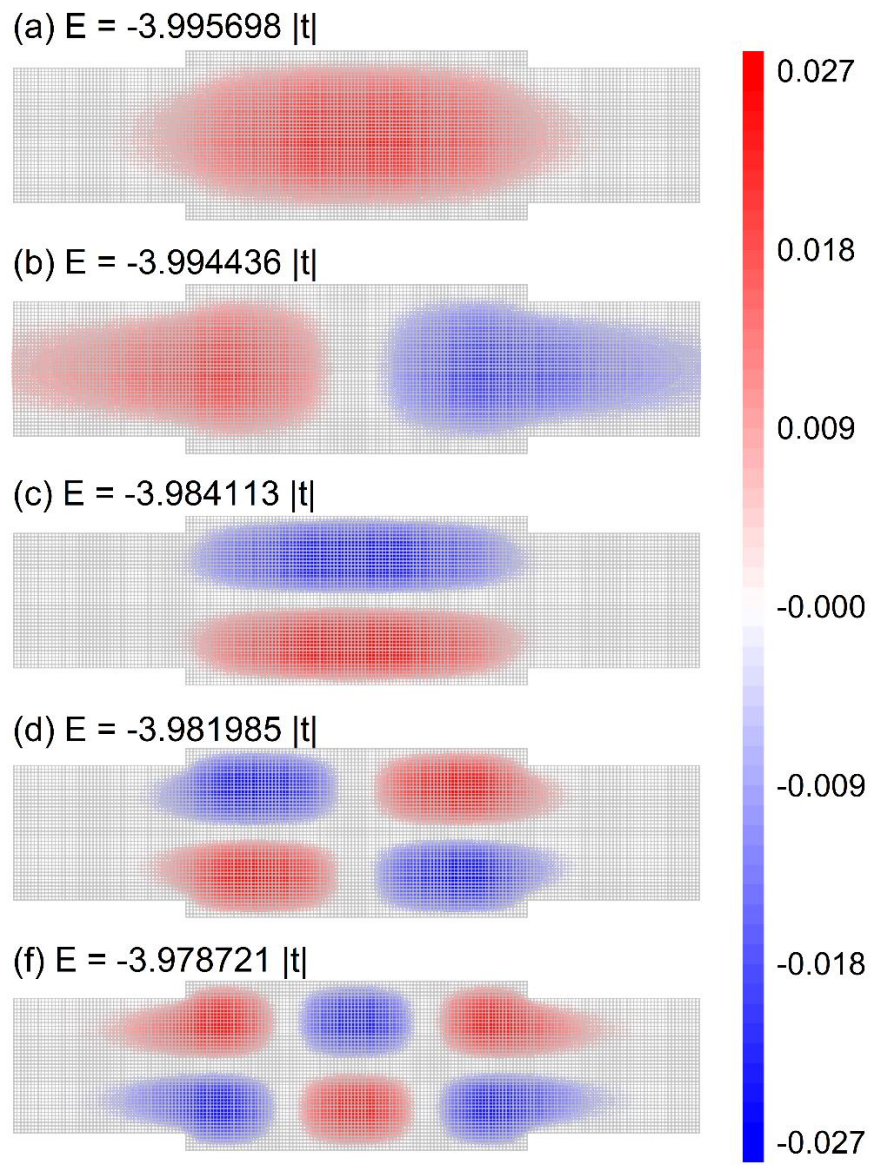

Fig. 3. First five bound-state eigenfunctions of the system in Fig. 1(a) with $N = 50$, $N_L = 40$ and $M = 100$.

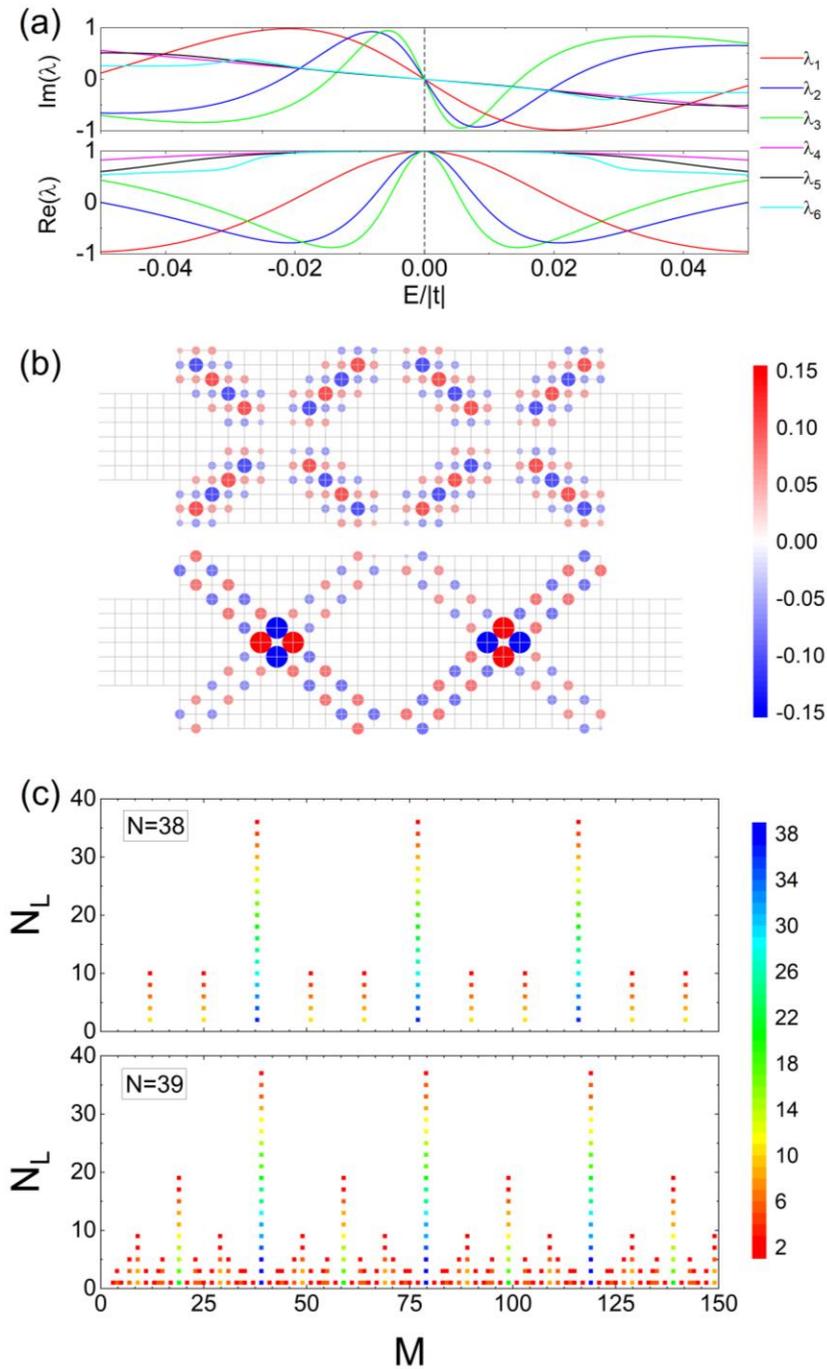

Fig 4. (a) Real (bottom) and imaginary (top) parts of the eigenvalues corresponding to a multiple degenerate bound state in a square lattice nanoribbon with a wider section. Only eigenvalues that become one at $E=0$ are shown. Part (b) shows two of the corresponding wavefunctions. (c) Number of bound states at $E=0$ for different values of $N$, $N_L$ and $M$.

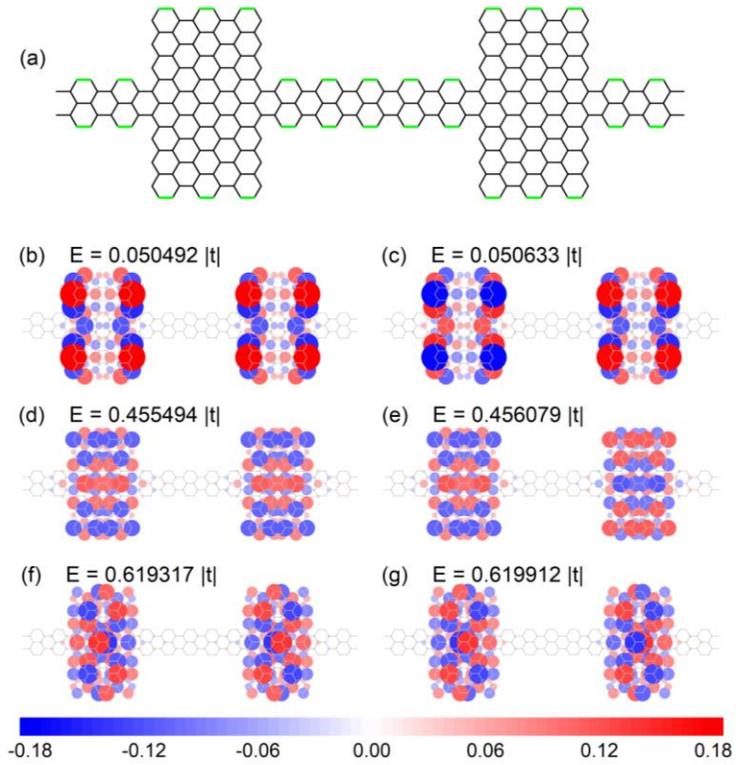

Fig. 5: (a) AGNR with two wider sections. The hopping parameter for the border dimers (green) was taken as $1.12t$ with $t$ being the parameter for the rest of cases (black). (b-g) Bound-state eigenfunctions of structure (a) with positive energies.

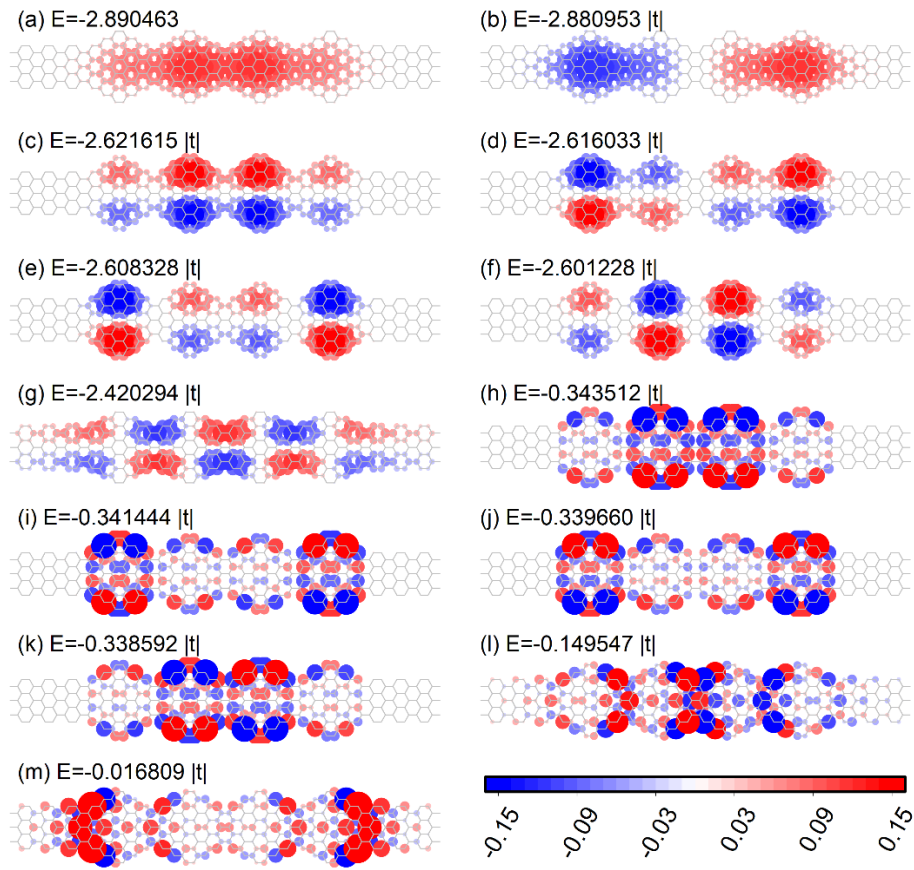

Fig. 6: First thirteen bound-state eigenfunctions for an AGNR with continuously varying width (gray lines).